\documentclass[showpacs,aps,graphicx]{revtex4}%,twocolumn
\usepackage{amssymb}
\usepackage{amsmath}%preprint,
\usepackage{graphicx}
\usepackage{multirow}

\begin{document}

%\begin{CJK*}{GBK}{song} %显示中文

%Robust-fidelity

\title{Self-error-corrected hyperparallel photonic quantum computation working with both the polarization and the spatial-mode degrees of freedom\footnote{Published in Opt. Express \textbf{26}, 23333--23346 (2018).}}

\author{Guan-Yu Wang,$^{1}$ Tao Li,$^{2}$ Qing Ai,$^{1}$  and Fu-Guo Deng$^{1,}$\footnote{Corresponding author:
fgdeng@bnu.edu.cn} }

\address{$^{1}$ Department of Physics, Applied Optics Beijing Area Major Laboratory,
Beijing Normal University, Beijing 100875, China \\
$^{2}$ School of Science, Nanjing University of Science and Technology, Nanjing 210094, China}

\date{\today }

\begin{abstract}
Usually, the hyperparallel quantum computation can speed up quantum computing, reduce the quantum resource consumed largely, resist to noise, and simplify the storage of quantum information. Here, we present the first scheme for the self-error-corrected hyperparallel photonic quantum computation working with both the polarization and the spatial-mode degrees of freedom of photon systems simultaneously. It can prevent bit-flip errors from happening with an imperfect nonlinear interaction in the nearly realistic condition. We give the way to design the universal hyperparallel photonic quantum controlled-NOT (CNOT) gate on a two-photon system, resorting to the nonlinear interaction between the circularly polarized photon and the electron spin in the quantum dot  in a double-sided microcavity system, by taking the imperfect interaction in the nearly realistic condition into account. Its self-error-corrected pattern prevents the bit-flip errors from happening in the hyperparallel quantum CNOT gate, guarantees the robust fidelity, and relaxes the requirement for its experiment. Meanwhile, this scheme works in a failure-heralded way. Also, we generalize this approach to achieve the self-error-corrected hyperparallel quantum CNOT$^N$ gate working on a multiple-photon system. These good features make this scheme more useful in the photonic quantum computation and quantum communication in the future.
\end{abstract}

 \maketitle

\section{Introduction}\label{sec1}

Quantum computation works as a pattern of parallel computing and it
has a stronger capacity in information processing. It can speed up
factorizing a large number \cite{Shor94} and searching data
\cite{Grover97,Long01}. The ultimate aim of quantum computation is
the realization of quantum computer and the key element of a quantum
computer is the quantum controlled-NOT (CNOT) gate (or its
equivalency, the quantum controlled-phase gate). Many physical
architectures have been proposed to implement quantum computation,
such as photons
\cite{pgate1,pgate2,pgate3,pgate4,pgate5,pgate8,pgate9,pgate10,pgate11,pgate12},
atoms \cite{atgate1,atgate2,atgate3,atgate4,atgate6,atgate7,atgate8,atgate9,atgate10},
quantum dots \cite{qdgate1,qdgate3,qdgate5,qdgate6,qdgate7,qdgate4}, diamond nitrogen-vacancy defect
centers \cite{nvgate1,nvgate2}, nuclear magnetic resonance
\cite{nmrgate1,nmrgate2,nmrgate3,nmrgate4,nmrgate5,nmrgate6,nmrgate7},
and so on. Photon is an idea information carrier as it has a high
transmission speed, a weak interaction with its environment, and a
low cost for its preparation. Moreover, a photon system can have multiple degrees of freedom (DOFs), which can be used in various quantum information processing tasks \cite{pqip1,pqip2,pqip3,pqip4,pqip5}. A state of a photon system being
simultaneously entangled in several DOFs is referred to as a hyperentanglement, and its generation has been extensively researched both theoretically and experimentally \cite{hbsg1,hbsg2,hbsg3,hbsg4,hbsg5,hbsg6,hbsg7,hbsg8,hbsg9,hbsg10,hbsg11}. The hyperentangled states of the photon systems can improve
both the channel capacity and the security of quantum communication
largely. The hyperparallel quantum computation is accomplished with
the hyperentangled states of the photon systems. The hyperparallel
quantum computation can achieve the full potential of the parallel
computing in quantum computation, and it can speed up quantum
computing, save quantum resource, resist to noise, and will be
useful for easy storage in the following process.

The first protocol for the hyperparallel quantum CNOT gate was proposed by Ren
\emph{et al}. \cite{hyper1} in 2013. In their protocol, a
deterministic hyper-CNOT gate is constructed and it
operates in both the spatial-mode and the polarization DOFs for a photon
pair simultaneously. Up to now, several hyperparallel quantum gate
protocols have been proposed, including the hyperparallel CNOT gate on a two-photon system with two DOFs and the hyperparallel Toffoli gate on a three-photon system with two DOFs \cite{hyper2,hyper3,hyper4,hyper5}. A deterministic (hyperparallel)
photonic quantum gate can be completed with nonlinearity
interaction, which can be provided by a matter qubit such as an
artificial atom trapped in a microcavity. The electron spin in a
GaAs-based or InAs-based charged quantum dot (QD) trapped in a double-sided
microcavity is an attractive matter qubit \cite{qdgate6,huqd1}.
Utilizing the technique of spin echo, the electron spin coherence
time of a charged QD can be maintained for more than 3 $\mu s$
\cite{qdtime1,qdtime2}, and the electron spin relaxation time can be
longer ($\sim$ $ms$) \cite{qdtime3,qdtime4}. The techniques of fast
preparing the superposition states of an electron spin in a singly charged
QD \cite{qdpre1,qdpre2}, fast manipulating the electron spin in a
charged QD \cite{qdmani1,qdmani2,qdmani3,qdmani4}, and detecting the
state of the electron spin in a charged QD \cite{qddete} have been
realized. Moreover, it is easy to embed a charged QD into a
solid-state microcavity.

In the ideal case, the QD-microcavity system can provide a perfect
nonlinear interaction which can be used to construct a
unity-fidelity deterministic (hyperparallel) photonic quantum gate.
When it turns into the realistic condition, the nonlinear interaction is not perfect, which will reduce the fidelity of the quantum gate.
Fidelity, which is the description of the quality of the quantum
gate, is defined as $F=|\langle \Psi_r|\Psi _i\rangle|^2$,
where $|\Psi_r\rangle$ and $|\Psi_i\rangle$ are the realistic
final state and the ideal final state, respectively. In the realistic condition, $|\Psi_r\rangle\neq|\Psi_i\rangle$, which means there would be some errors in the final result of a quantum gate and leads to a non-unity fidelity. For a quantum CNOT gate, there would be bit-flip errors in the realistic condition and the fidelity would be non-unity. Specifically speaking, for a quantum CNOT gate,
$|\Psi_0\rangle=(a_1|0\rangle_c+a_2|1\rangle_c)(b_1|0\rangle_t+b_2|1\rangle_t)$ represents the initial state of the system, where the subscripts $c$ and $t$ respectively represent the control qubit and the target qubit, and the coefficients satisfy $|a_1|^2+|a_2|^2=1$ and $|b_1|^2+|b_2|^2=1$. $|\Psi_i\rangle=a_1|0\rangle_c(b_1|0\rangle_t+b_2|1\rangle_t)
+a_2|1\rangle_c(b_2|0\rangle_t+b_1|1\rangle_t)$ is the ideal final state after a quantum CNOT gate, which means the state of the target qubit is flipped when the control qubit is in the state $|1\rangle_c$, while it would not be changed when the control qubit is in the state $|0\rangle_c$. However, in the
realistic condition, after the evolution of the system, the final
state might be
$|\Psi_r\rangle=a'_1|0\rangle_c(b_1|0\rangle_t+b_2|1\rangle_t)
+a'_2|1\rangle_c(b_2|0\rangle_t+b_1|1\rangle_t)
+a'_3|0\rangle_c(b_2|0\rangle_t+b_1|1\rangle_t)
+a'_4|1\rangle_c(b_1|0\rangle_t+b_2|1\rangle_t)$, which
means there is a possibility that the state of the target qubit is
not flipped when the control qubit is in the state $|1\rangle_c$ and
the state of the target qubit is flipped when the control qubit is in the state
$|0\rangle_c$. The coefficients $a'_1$, $a'_2$, $a'_3$, and
$a'_4$ and the fidelity of the quantum CNOT gate are affected by the
parameters of the photon and the QD-microcavity system. That is to say, in a quantum CNOT gate together with a hyperparallel quantum CNOT gate, bit-flip errors can happen in the realistic condition, which would be reflected by the non-unity fidelity. As the importance of improving the fidelity of a quantum gate, the research in quantum gates with a robust fidelity has attracted much attention and several interesting schemes of quantum gates for atom systems \cite{atgate7,atgate8,atgate9} and QD systems \cite{qdgate4} with robust fidelities have been proposed.

In this paper, we give an original approach to construct the universal self-error-corrected hyperparallel photonic quantum CNOT gate working on a two-photon system in both the polarization and the spatial-mode DOFs.
In this gate, the state of one photon in the polarization and the spatial-mode DOFs controls the state of the other photon in the polarization and the spatial-mode DOFs simultaneously, which is equal to two identical quantum CNOT gates operating simultaneously on the systems in one DOF. It can speed up quantum computing, reduce the quantum resource consumed largely, resist to noise, and simplify the storage of quantum information in a practical application. More interestingly, it can prevent the bit-flip errors arising from  the nearly realistic imperfect nonlinear interaction of the QD-microcavity system. Its self-error-corrected pattern prevents bit-flip errors from happening in the gate, guarantees its fidelity  robust, and relaxes its requirement for the realistic parameters. This self-error-corrected hyperparallel photonic quantum CNOT gate also works in a failure-heralded way. Meanwhile, we generalize this approach of the self-error-corrected hyperparallel photonic quantum CNOT gate working on a two-photon system to achieve the self-error-corrected hyperparallel photonic quantum CNOT$^N$ gate working on a multiple-photon system, which also works in a failure-heralded way. These good features make this scheme more useful in the hyperparallel photonic quantum computation and the multiple-photon hyperentangled-state generation for quantum communication in the future.

\begin{figure}[ht!]
\centering
\includegraphics[width=10cm,angle=0]{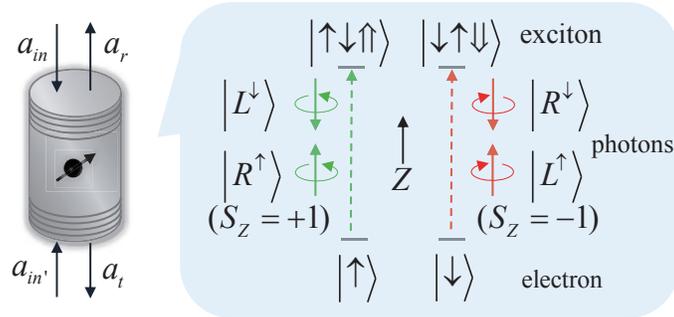}
\caption{ Schematic diagrams for a singly charged QD in a
double-sided microcavity,  the relevant energy levels, and the
optical transitions. Z represents the quantization axis (microcavity Z
axis). $|R^\uparrow\rangle$ ($|R^\downarrow\rangle$) and
$|L^\uparrow\rangle$ ($|L^\downarrow\rangle$) represent the
right-circularly polarized photon and the left-circularly polarized
photon propagating along (against) the normal direction of the
microcavity Z axis, respectively.} \label{fig1}
\end{figure}

\section{Nonlinear interaction between a circularly polarized photon
and a QD in a double-sided microcavity} \label{sec2}

A singly charged electron In(Ga)As QD or a GaAs interface QD is set
at the wave loop of a double-sided microcavity. The reflection
coefficients of the bottom and the top distributed Bragg reflectors
of the double-sided microcavity are the same. An excess electron is
injected into the QD. After the optical excitation by a circularly
polarized photon, an exciton $X^-$ consisting of two electrons
bound to one heavy hole with negative charges is created. According
to Pauli's exclusion principle, the circularly polarized photon
$|L^\downarrow\rangle$ or $|R^\uparrow\rangle$ ($S_z=+1$) can
only realize the transition from $|\uparrow\rangle$ to
$|\uparrow\downarrow\Uparrow\rangle$, and the circularly polarized
photon $|R^\downarrow\rangle$ or $|L^\uparrow\rangle$
($S_z=-1$) can only realize the transition from
$|\downarrow\rangle$ to $|\downarrow\uparrow\Downarrow\rangle$, as
shown in Fig.~\ref{fig1}.

The dipole interaction process can be represented by Heisenberg
equations of motion for the annihilation operator $\hat{a}$ of
the microcavity mode and the dipole operator $\hat{\sigma}_-$ of
the exciton $X^-$. The Heisenberg equations in the interaction
picture and the input-output relationships are described as
\cite{Walls94}
\begin{eqnarray}\label{eq1}
\begin{split}
\frac{d\hat{a}}{dt}\;=\;&-[i(\omega_c-\omega)+\kappa
+\frac{\kappa_s}{2}]\hat{a}-g\hat{\sigma}_{-}-\sqrt{\kappa}(\hat{a}_{in'}+\hat{a}_{in})+\hat{H},\\
\frac{\hat{\sigma}_{-}}{dt}\;=\;&-[i(\omega_{X^-}-\omega)
+\frac{\gamma}{2}]\hat{\sigma}_- -g\hat{\sigma}_z\hat{a}+\hat{G},\\
\hat{a}_r\;=\;&\hat{a}_{in}+\sqrt{\kappa}\hat{a},\\
\hat{a}_t\;=\;&\hat{a}_{in'}+\sqrt{\kappa}\hat{a}.
\end{split}
\end{eqnarray}
Here $g$ is the coupling strength between the $X^-$ and the microcavity. $\omega$, $\omega_{c}$, and $\omega_{X^{-}}$ are frequencies of
the photon, the microcavity, and $X^-$ transition, respectively.
$\gamma$, $\kappa$, and $\kappa_s$ represent $X^-$ dipole decay
rate, the microcavity decay rate, and the microcavity leaky rate, respectively.
$\hat{H}$ and $\hat{G}$ are noise operators.
$\hat{a}_{in}$, $\hat{a}_{in'}$, $\hat{a}_r$, and
$\hat{a}_t$ are the input and output field operators. In the
weak excitation approximation ($\langle\hat{\sigma}_z\rangle=-1$),
after the nonlinear interaction between the circularly polarized
photon and the QD-microcavity system, the reflection coefficient
$r$ and the transmission coefficient $t$ can be
described by \cite{qdgate6,anjunhong}
\begin{eqnarray}\label{eq2}
r\;=\;1+t,\;\;\;\;\;\;\;t\;=\;-\frac{\kappa[i(\omega_{X^{-}}-\omega)
+\frac{\gamma}{2}]}{[i(\omega_{X^{-}}-\omega)
+\frac{\gamma}{2}][i(\omega_c-\omega)+\kappa+\frac{\kappa_s}{2}]+g^2}.
\end{eqnarray}
If the circularly polarized photon interacts with a cold microcavity,
that is $g=0$, the reflection $r_0$ and the transmission
$t_0$ coefficients can be written as
\begin{eqnarray}\label{eq3}
r_0\;=\;1+t_0,\;\;\;\;\;\;\;t_0\;=\;-\frac{\kappa}{i(\omega_c-\omega)+\kappa+\frac{\kappa_s}{2}}.
\end{eqnarray}
The evolution rules for the nonlinear interaction between the circularly
polarized photon and the QD in a double-sided microcavity in the
realistic condition can be described as \cite{qdgate6}
\begin{eqnarray}    \label{eq4}
\begin{split}
&|R^\uparrow\uparrow\rangle \;\rightarrow
\;r|L^\downarrow\uparrow\rangle + t|R^\uparrow\uparrow\rangle,
\;\;\;\;\;\;\;\;\;\;\;
|L^\downarrow\uparrow\rangle \; \rightarrow \;r|R^\uparrow\uparrow\rangle+t|L^\downarrow\uparrow\rangle,\\
&|R^\downarrow\uparrow\rangle \; \rightarrow \;
t_0|R^\downarrow\uparrow\rangle
+r_0|L^\uparrow\uparrow\rangle, \;\;\;\;\;\;\;\;
|L^\uparrow\uparrow\rangle \; \rightarrow t_0|L^\uparrow\uparrow\rangle +r_0|R^\downarrow\uparrow\rangle,\\
&|R^\downarrow\downarrow\rangle \; \rightarrow
r|L^\uparrow\downarrow\rangle + t|R^\downarrow\downarrow\rangle,
\;\;\;\;\;\;\;\;\;\;\;\;
|L^\uparrow\downarrow\rangle \;\rightarrow \;r|R^\downarrow\downarrow\rangle + t |L^\uparrow\downarrow\rangle,\\
&|R^\uparrow\downarrow\rangle \;\rightarrow\;
t_0|R^\uparrow\downarrow\rangle+r_0|L^\downarrow\downarrow\rangle,
\;\;\;\;\;\;\;\; |L^\downarrow\downarrow\rangle \;\rightarrow\;
t_0 |L^\downarrow\downarrow\rangle
+r_0|R^\uparrow\downarrow\rangle.
\end{split}
\end{eqnarray}

\begin{figure}[ht!]
\centering
\includegraphics[width=14cm,angle=0]{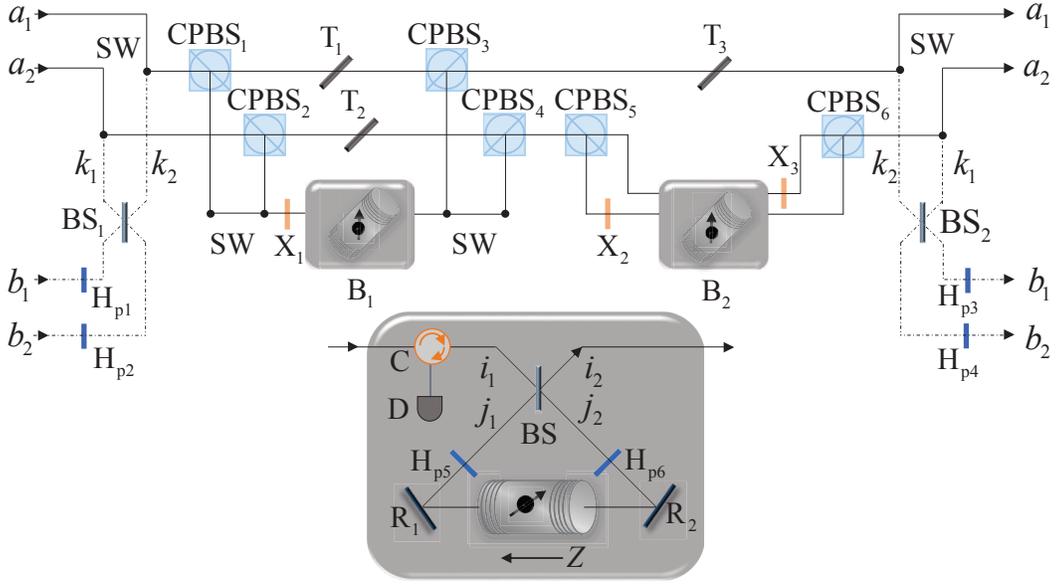}
\caption{ Schematic diagram for hyperparallel
photonic quantum CNOT gate with a self-error-corrected pattern.
CPBS$_i(i=1\sim6)$ is a circularly polarizing beam splitter which
transmits the photon in the right-circular polarization $|R\rangle$
and reflects the photon in the left-circular polarization
$|L\rangle$, respectively. BS$_i(i=1,2)$ is a $50:50$ beam splitter which performs the spatial-mode Hadamard operation
$[|b_1\rangle\rightarrow\frac{1}{\sqrt{2}}(|c_1\rangle+|c_2\rangle),
|b_2\rangle\rightarrow\frac{1}{\sqrt{2}}(|c_1\rangle-|c_2\rangle)]$
on the photon. H$_{pi}(i=1\sim4$) is a half-wave plate which
performs the polarization Hadamard operation
$[|R\rangle\rightarrow\frac{1}{\sqrt{2}}(|R\rangle+|L\rangle),
|L\rangle\rightarrow\frac{1}{\sqrt{2}}(|R\rangle-|L\rangle)]$ on the
photon. X$_i(i=1\sim3)$ is a half-wave plate which performs a
bit-flip operation $[|R\rangle\rightarrow|L\rangle,|L\rangle\rightarrow|R\rangle]$ on the photon. T$_i(i=1\sim3)$ is a partially transmitted mirror with the
transmission coefficient $T$. SW is an optical switch which couples different photons into and out of the circuit of the photonic quantum gate sequentially and couples different spatial modes of a photon into and out of the basic
block sequentially. B$_i(i=1,2)$ represents the basic block consisting of the QD$_i(i=1,2)$-microcavity and the schematic diagram for the basic block is shown in the inset, in which C is an optical circulator which can route the photon into an appropriate path. BS is a $50:50$ beam
splitter. H$_{pi}(i=5,6)$ is a half-wave plate and completes the polarization Hadamard operation. R$_i(i=1,2)$ is a fully reflective mirror. D is a single-photon detector. } \label{fig2}
\end{figure}

\section{Self-error-corrected hyperparallel photonic quantum CNOT
gate} \label{sec3}

The hyperparallel photonic quantum CNOT gate working on a two-photon system
in both the polarization and the spatial-mode DOFs, completes
the task that when the polarization of photon $a$ (the control
qubit) is in the state $|L\rangle$, a bit flip takes place on the
state in the polarization DOF of photon $b$ (the target qubit), and
simultaneously when the spatial mode of photon $a$ is in the state
$|a_2\rangle$, a bit flip takes place on the state in the
spatial-mode DOF of photon $b$. The schematic diagram for our
self-error-corrected hyperparallel photonic quantum CNOT gate is shown in Fig.~\ref{fig2}. The QD-microcavity system is set in a basic block and the circularly polarized photon interacts with the QD-microcavity system through the
basic block. In this way, the hyperparallel photonic quantum CNOT gate is constructed with a self-error-corrected pattern and works in a failure-heralded way.

The schematic diagram for the basic block is shown in the inset of
Fig.~\ref{fig2}. The electron spin in the QD in the microcavity is
initially prepared in the state
$|\varphi_+\rangle=\frac{1}{\sqrt{2}}(|\uparrow\rangle+|\downarrow\rangle)$
and a right-circularly polarized photon $|R\rangle$ is injected into
the basic block from the input path.  After the photon passing
through BS and H$_{pi}(i=5,6)$, it interacts with the QD-microcavity
system. The rules of the nonlinear interaction between the photon
and the QD-microcavity system in the realistic condition is governed by
Eq.~(\ref{eq4}). And at last the photon passes through H$_{pi}(i=5,6)$ and BS
again. The final state of the system consisting of
the photon and the electron spin in the QD becomes
\begin{eqnarray}\label{eq5}
|\Phi\rangle_1\;=\;D|R\rangle|i_1\rangle|\varphi_+\rangle
+T|L\rangle|i_2\rangle|\varphi_-\rangle.
\end{eqnarray}
Similarly, when the electron spin in the QD in the microcavity is initially
prepared in the state
$|\varphi_-\rangle=\frac{1}{\sqrt{2}}(|\uparrow\rangle-|\downarrow\rangle)$,
after a right-circularly polarized photon $|R\rangle$ passing
through the basic block from the input path, the final state
of the system becomes
\begin{eqnarray}\label{eq6}
\begin{split}
|\Phi\rangle_2\;=\;D|R\rangle|i_1\rangle|\varphi_-\rangle
+T|L\rangle|i_2\rangle|\varphi_+\rangle.
\end{split}
\end{eqnarray}
$D=\frac{1}{2}(t+r+t_0 + r_0)$ and
$T=\frac{1}{2}(t+r-t_0 - r_0)$ in Eqs.~(\ref{eq5}) and (\ref{eq6})
are the reflection coefficient and the transmission coefficient for
the basic block. Equations.~(\ref{eq5}) and (\ref{eq6}) describe the rules
of the interaction between the photon and the basic block. If the
photon is reflected from the basic block with the probability of
$|D|^2$, the polarization of the photon and the state of  the
electron spin in the QD would not be changed. The reflected photon
would be detected by the single-photon detector and the following
process would be terminated. If the photon is transmitted from the
basic block with the probability of $|T|^2$, the polarization of
the photon would be changed and the state of the electron spin in
the QD would be changed as
$[|\uparrow\rangle\rightarrow|\uparrow\rangle,
|\downarrow\rangle\rightarrow-|\downarrow\rangle]$.  The transmitted
photon will be used in the following process and the system
consisting of the photon and the electron spin in the QD has an integrated coefficient $T$, which can guarantee the fidelity unity in the nearly realistic condition.

Based on the rules of the interaction between a right-circularly
polarized photon $|R\rangle$ and a basic block, a self-error-corrected hyperparallel photonic quantum CNOT gate working in a failure-heralded way can be constructed. The detail principle can be described as follows in step by step.

Initially, the two electron spins in QD$_1$ and QD$_2$ are prepared in
the states $|\varphi_+\rangle_1$ and $|\varphi_+\rangle_2$,
and the two photons $a$ and $b$ are in the states
$|\phi\rangle_a=(\alpha_1|R\rangle
+\alpha_2|L\rangle)_a(\beta_1|a_1\rangle+\beta_2|a_2\rangle)$
and $|\phi\rangle_b=(\gamma_1|R\rangle
+\gamma_2|L\rangle)_b(\delta_1|b_1\rangle+\delta_2|b_2\rangle)$,
respectively. Here, $|\alpha_1|^2+|\alpha_2|^2=1$, $|\beta_1|^2+|\beta_2|^2=1$, $|\gamma_1|^2+|\gamma_2|^2=1$, and $|\delta_1|^2+|\delta_2|^2=1$.

First, photon $a$ is injected into the circuit from the input path
$a_i(i=1,2)$. After photon $a$ passes through CPBS$_1$ and CPBS$_2$, in both of the two spatial modes $|a_1\rangle$ and $|a_2\rangle$, the wave packet
in the left-circular polarization will pass through X$_1$ and
the first basic block B$_1$, and the wave
packet in the right-circular polarization will pass through
T$_1$ and T$_2$. If there is a click of the single-photon
detector in B$_1$, the process of the self-error-corrected hyperparallel photonic quantum CNOT gate fails and  it is terminated. If the single-photon
detector in B$_1$ does not click, the whole
process will continue. The wave packet in the left-circular
polarization and the wave packet in the right-circular
polarization would reunion after passing through CPBS$_{3,4}$. In
this time, the state of the hybrid system consisting of two photons
$a$ and $b$ and two electron spins in QD$_1$ and QD$_2$ is changed from
$|\Psi\rangle_0=|\varphi_+\rangle_1|\varphi_+\rangle_2|\phi\rangle_a|\phi\rangle_b$
to $|\Psi\rangle_1$. Here
\begin{eqnarray}\label{eq7}
|\Psi\rangle_1\;=\;T(\beta_1|a_1\rangle
+\beta_2|a_2\rangle)(\alpha_1|R\rangle_a|\varphi_+\rangle_1+\alpha_2|L\rangle_a|\varphi_-\rangle_1)\otimes|\phi\rangle_b|\varphi_+\rangle_2.
\end{eqnarray}
Subsequently, for the wave packet of photon $a$ in the spatial mode $|a_1\rangle$, it transmits through T$_3$. For the wave packet of photon $a$ in the spatial mode $|a_2\rangle$, after it passes through CPBS$_5$, the wave packet in the left-circular polarization transmits through X$_2$ and the second basic block B$_2$ in sequence, and the wave packet in the right-circular polarization transmits through the second basic block B$_2$ and X$_3$ in sequence. Similarly, if there is a click of the single-photon detector in B$_2$, the process of the self-error-corrected hyperparallel photonic quantum CNOT gate  fails and it is terminated. If
there is no click of the single-photon detector in B$_2$, the two wave packets of different polarizations in the spatial mode $|a_2\rangle$ reunion
at CPBS$_6$, and the photon $a$ comes out of the quantum circuit
from path $a_i(i=1,2)$. And then Hadamard operations
$[|\varphi_+\rangle\rightarrow|\uparrow\rangle,
|\varphi_-\rangle\rightarrow|\downarrow\rangle]$ are respectively performed on the electron spins in QD$_1$ and QD$_2$. The state of the hybrid system is changed
into
\begin{eqnarray}\label{eq8}
|\Psi\rangle_2\;=\;T^2(\alpha_1|R\rangle_a|\uparrow\rangle_1
+\alpha_2|L\rangle_a|\downarrow\rangle_1)(\beta_1|a_1\rangle|\uparrow\rangle_2+\beta_2|a_2\rangle|\downarrow\rangle_2)
\otimes|\phi\rangle_b.
\end{eqnarray}

Second, photon $b$ is injected into the circuit from the input path
$b_i(i=1,2)$. After photon $b$ passes through H$_{p1,2}$
and BS$_1$, the state of the hybrid system becomes
\begin{eqnarray}\label{eq9}
\begin{split}
|\Psi\rangle_3\;=\;T^2(\alpha_1|R\rangle_a|\uparrow\rangle_1
+\alpha_2|L\rangle_a|\downarrow\rangle_1)(\beta_1|a_1\rangle|\uparrow\rangle_2+\beta_2|a_2\rangle|\downarrow\rangle_2)
\otimes[(\gamma'_1|R\rangle+\gamma'_2|L\rangle)_b(\delta'_1|b_1\rangle
+\delta'_2|b_2\rangle)].
\end{split}
\end{eqnarray}
Here,
$\gamma'_{1}=\frac{1}{\sqrt{2}}(\gamma_1+\gamma_2),\gamma'_2
=\frac{1}{\sqrt{2}}(\gamma_1-\gamma_2),\delta'_1
=\frac{1}{\sqrt{2}}(\delta_1+\delta_2),\delta'_2
=\frac{1}{\sqrt{2}}(\delta_1-\delta_2)$. Subsequently, photon
$b$ passes through the circuit consisting of the optical elements
and two basic blocks. Similarly, if there is a click of the single-photon
detector in B$_1$ or B$_2$, the process of the self-error-corrected hyperparallel photonic quantum CNOT gate  fails and it is terminated. If there is no click, the whole process will continue. The polarization Hadamard and spatial-mode
Hadamard operations are performed again on photon $b$ by
H$_{P3,4}$ and BS$_2$, respectively. Hadamard operations are
performed either on the two electron spins in QD$_1$ and QD$_2$
again. In this time, the state of the system consisting of two photons $a$ and $b$
and two electron spins in QD$_1$ and QD$_2$ is changed into
\begin{eqnarray}\label{eq10}
%\begin{split}
|\Psi\rangle_4=\!\!\!\!\!\!&&\frac{T^4}{2}[\alpha_1|R\rangle_a(\gamma_1|R\rangle
+\gamma_2|L\rangle)_b+\alpha_2|L\rangle_a(\gamma_2|R\rangle+\gamma_1|L\rangle)_b][\beta_1|a_1\rangle(\delta_1|b_1\rangle
+\delta_2|b_2\rangle)+\beta_2|a_2\rangle(\delta_2|b_1\rangle
+\delta_1|b_2\rangle)]|\uparrow\uparrow\rangle_{12}\nonumber\\
&&+\frac{T^4}{2}[\alpha_1|R\rangle_a(\gamma_1|R\rangle
+\gamma_2|L\rangle)_b-\alpha_2|L\rangle_a(\gamma_2|R\rangle+\gamma_1|L\rangle)_b][\beta_1|a_1\rangle(\delta_1|b_1\rangle
+\delta_2|b_2\rangle)+\beta_2|a_2\rangle(\delta_2|b_1\rangle
+\delta_1|b_2\rangle)]|\downarrow\uparrow\rangle_{12}\nonumber\\
&&+\frac{T^4}{2}[\alpha_1|R\rangle_a(\gamma_1|R\rangle
+\gamma_2|L\rangle)_b+\alpha_2|L\rangle_a(\gamma_2|R\rangle+\gamma_1|L\rangle)_b][\beta_1|a_1\rangle(\delta_1|b_1\rangle
+\delta_2|b_2\rangle)-\beta_2|a_2\rangle(\delta_2|b_1\rangle
+\delta_1|b_2\rangle)]|\uparrow\downarrow\rangle_{12}\nonumber\\
&&+\frac{T^4}{2}[\alpha_1|R\rangle_a(\gamma_1|R\rangle
+\gamma_2|L\rangle)_b-\alpha_2|L\rangle_a(\gamma_2|R\rangle+\gamma_1|L\rangle)_b][\beta_1|a_1\rangle(\delta_1|b_1\rangle
+\delta_2|b_2\rangle)-\beta_2|a_2\rangle(\delta_2|b_1\rangle
+\delta_1|b_2\rangle)]|\downarrow\downarrow\rangle_{12}.\nonumber\\
%\end{split}
\end{eqnarray}

At last, the electron spins in QD$_1$ and QD$_2$ are measured in
the orthogonal basis $[|\uparrow\rangle,|\downarrow\rangle]$. If the
electron spin in QD$_1$ is in the state $|\downarrow\rangle_1$,
an additional phase-flip operation
$|L\rangle_{a}\rightarrow-|L\rangle_a$ is performed on photon $a$,
and if the electron spin in QD$_2$ is in the state
$|\downarrow\rangle_2$, an additional phase-flip operation
$|a_2\rangle\rightarrow-|a_2\rangle$ is performed on photon $a$.
Conditioned on the results of the measurement on the QD$_1$ and QD$_2$, the two-photon system, up to a single-photon rotation, is collapsed into the desired state with a success probability $T^{8}/4$ as follows
\begin{eqnarray}\label{eq11}
|\Psi\rangle_{p}=[\alpha_1|R\rangle_a(\gamma_1|R\rangle
+\gamma_2|L\rangle)_b+\alpha_2|L\rangle_a(\gamma_2|R\rangle+\gamma_1|L\rangle)_b]
\otimes[\beta_1|a_1\rangle(\delta_1|b_1\rangle
+\delta_2|b_2\rangle)+\beta_2|a_2\rangle(\delta_2|b_1\rangle+\delta_1|b_2\rangle)].
\end{eqnarray}

From Eq.~(\ref{eq11}), one can see that there is a bit flip of the state on the
polarization DOF for photon $b$ when the polarization of photon $a$
is in the state $|L\rangle_a$, and there is a bit flip of the
state on the spatial-mode DOF for photon $b$ when the spatial mode
of photon $a$ is in the state $|a_2\rangle$. Meanwhile, from Eq.~(\ref{eq11}), one can see that in the nearly realistic condition with imperfect nonlinear interaction, bit-flip errors do not happen in our hyperparallel photonic quantum CNOT gate, which means it works in the self-error-corrected pattern. Furthermore, from Eq.~(\ref{eq10}), one can see that $\frac{T^4}{2}$ is an integrated
coefficient for the state of the two-photon system and the two electron spins in the QDs. Transmission coefficient $T$ for the basic block is the function of transmission coefficient $t~(t_0)$ and reflection coefficient $r~(r_0)$ which are affected by the experimental parameters $g$, $\omega$, $\omega_{c}$, $\omega_{X^{-}}$, $\gamma$, $\kappa$, and $\kappa_s$. That is to say, these experimental parameters would not affect the fidelity of the hyperparallel photonic quantum CNOT gate, which relaxes the requirement for experiment. Overall, the schematic diagram shown in Fig.~\ref{fig2} completes a self-error-corrected hyperparallel photonic quantum CNOT gate in the nearly realistic condition, which eliminates bit-flip errors, guarantees a robust fidelity, and relaxes the requirement for experiment, and its failure is heralded by the single-photon detectors.

\section{Self-error-corrected hyperparallel photonic quantum CNOT$^N$ gate} \label{sec4}

The previous approach completes the self-error-corrected hyperparallel quantum CNOT gate working on a two-photon system, which can be generalized to achieve the self-error-corrected hyperparallel quantum CNOT$^N$ gate working on a multiple-photon system. The self-error-corrected hyperparallel photonic quantum CNOT$^N$ gate working on a multiple-photon system is completed with two basic blocks consisting of two QD-microcavity systems as auxiliaries, which is the same as the self-error-corrected hyperparallel quantum CNOT gate working on a two-photon system. Initially, the two electron spins in the two QDs are prepared in the states $|\varphi_+\rangle_1$ and $|\varphi_+\rangle_2$. The control photon $a$ is in the state $|\phi\rangle_a=(\alpha_1|R\rangle+\alpha_2|L\rangle)_a(\beta_1|a_1\rangle+\beta_2|a\rangle)$ with $|\alpha_1|^2+|\alpha_2|^2=1$ and $|\beta_1|^2+|\beta_2|^2=1$. The $N$ target photons $b^n(n=1,2,...,N)$ are in the states $|\phi\rangle_{b^{n}}=(\gamma_1^n|R\rangle
+\gamma_2^n|L\rangle)_{b^{n}}(\delta_1^n|b_1\rangle+\delta_2^n|b_2\rangle)_{b^{n}}$ with $|\gamma_1^n|^2+|\gamma_2^n|^2=1$ and $|\delta_1^n|^2+|\delta_2^n|^2=1$.

The schematic diagram shown in Fig.~\ref{fig2} can also be used to achieve the self-error-corrected hyperparallel photonic quantum CNOT$^N$ gate. First, the control photon $a$ is injected into the circuit from input path $a_i(i=1,2)$. During photon $a$ passing through the circuit, if there is a click of the single-photon detector in the basic block B$_1$ or B$_2$, the process of self-error-corrected hyperparallel photonic quantum CNOT$^N$ gate fails and it is terminated. If there is no click of the single-photon detector in B$_1$ nor B$_2$, with Hadamard operations respectively performed on the electron spins in QD$_1$ and QD$_2$, the state of the hybrid system is changed into
\begin{eqnarray}\label{eq12}
|\Psi^N\rangle_1\;=\; T^2(\alpha_1|R\rangle_a|\uparrow\rangle_1
+\alpha_2|L\rangle_a|\downarrow\rangle_1)(\beta_1|a_1\rangle|\uparrow\rangle_2+\beta_2|a_2\rangle|\downarrow\rangle_2)
\otimes|\phi\rangle_{b^{1}}|\phi\rangle_{b^{2}}\cdot\cdot\cdot|\phi\rangle_{b^{N}}.
\end{eqnarray}

Second, the first target photon $b^1$ is injected into the circuit from the input path $b_i(i=1,2)$. Similarly, during photon $b^1$ passing through the circuit, if there is a click of the single-photon detector in the basic block B$_1$ or B$_2$ the process of self-error-corrected hyperparallel photonic quantum CNOT$^N$ gate fails and it is terminated. If there is no click of the single-photon detector in B$_1$ nor B$_2$, the state of the hybrid system evolves into
\begin{eqnarray}\label{eq13}
\begin{split}
|\Psi^N\rangle_2\;=\;&T^4[\alpha_1|R\rangle_a|\uparrow\rangle_1(\gamma_1^1|R\rangle
+\gamma_2^1|L\rangle)_{b^{1}}
+\alpha_2|L\rangle_a|\downarrow\rangle_1(\gamma_2^1|R\rangle
+\gamma_1^1|L\rangle)_{b^{1}}]\\
&[\beta_1|a_1\rangle|\uparrow\rangle_2(\delta_1^1|b_1\rangle+\delta_2^1|b_2\rangle)_{b^{1}}+\beta_2|a_2\rangle|\downarrow\rangle_2(\delta_2^1|b_1\rangle+\delta_1^1|b_2\rangle)_{b^{1}}]
\otimes|\phi\rangle_{b^2}\cdot\cdot\cdot|\phi\rangle_{b^{N}}.
\end{split}
\end{eqnarray}

Subsequently, the target photon from $b^2$ to $b^N$ is injected into the circuit from the input path $b_i(i=1,2)$ one by one. The process is similar to the case that the click of the single-photon detector in B$_1$ or B$_2$ heralded the failure of the self-error-corrected hyperparallel photonic quantum CNOT$^N$ gate. If there is no click of the single-photon detector in B$_1$ nor B$_2$, after the last target photon $b^N$ exits from the output path $b_i(i=1,2)$, the state of the hybrid system is changed into
\begin{eqnarray}\label{eq14}
\begin{split}
|\!\Psi^N\!\rangle_3\!\!=&T^{2(N+1)}[\alpha_1|R\rangle_a|\!\uparrow\rangle_1(\gamma_1^1|R\rangle\!+\!\gamma_2^1|L\rangle)_{b^{1}}(\gamma_1^2|R\rangle\!+\!\gamma_2^2|L\rangle)_{b^{2}}\!\cdot\cdot\cdot\!(\gamma_1^N|R\rangle\!+\!\gamma_2^N|L\rangle)_{b^{N}}\\
&+\alpha_2|L\rangle_a|\downarrow\rangle_1(\gamma_2^1|R\rangle+\gamma_1^1|L\rangle)_{b^{1}}(\gamma_2^2|R\rangle+\gamma_1^2|L\rangle)_{b^{2}}\cdot\cdot\cdot(\gamma_2^N|R\rangle+\gamma_1^N|L\rangle)_{b^{N}}]\\
&[\beta_1|a_1\rangle|\uparrow\rangle_2(\delta_1^1|b_1\rangle+\delta_2^1|b_2\rangle)_{b^{1}}(\delta_1^2|b_1\rangle+\delta_2^2|b_2\rangle)_{b^{2}}\cdot\cdot\cdot(\delta_1^N|b_1\rangle+\delta_2^N|b_2\rangle)_{b^{N}}\\
&+\beta_2|a_2\rangle|\downarrow\rangle_2(\delta_2^1|b_1\rangle+\delta_1^1|b_2\rangle)_{b^{1}}(\delta_2^2|b_1\rangle+\delta_1^2|b_2\rangle)_{b^{2}}\cdot\cdot\cdot(\delta_2^N|b_1\rangle+\delta_1^N|b_2\rangle)_{b^{N}}].
\end{split}
\end{eqnarray}

At last, Hadamard operations are performed on both the two electron spins in QD$_1$ and QD$_2$, and they are measured in the orthogonal basis $[|\uparrow\rangle,|\downarrow\rangle]$. If the electron spin in QD$_1$ and the electron spin in QD$_2$ are respectively in the state $|\uparrow\rangle_1$ and $|\uparrow\rangle_2$, the control photon $a$ and the $N$ target photons are in the state
\begin{eqnarray}\label{eq15}
\begin{split}
|\Psi^N\rangle_p\;=\;&[\alpha_1|R\rangle_a(\gamma_1^1|R\rangle+\gamma_2^1|L\rangle)_{b^{1}}(\gamma_1^2|R\rangle+\gamma_2^2|L\rangle)_{b^{2}}\cdot\cdot\cdot(\gamma_1^N|R\rangle+\gamma_2^N|L\rangle)_{b^{N}}\\
&+\alpha_2|L\rangle_a(\gamma_2^1|R\rangle+\gamma_1^1|L\rangle)_{b^{1}}(\gamma_2^2|R\rangle+\gamma_1^2|L\rangle)_{b^{2}}\cdot\cdot\cdot(\gamma_2^N|R\rangle+\gamma_1^N|L\rangle)_{b^{N}}]\\
&\otimes[\beta_1|a_1\rangle(\delta_1^1|b_1\rangle+\delta_2^1|b_2\rangle)_{b^{1}}(\delta_1^2|b_1\rangle+\delta_2^2|b_2\rangle)_{b^{2}}\cdot\cdot\cdot(\delta_1^N|b_1\rangle+\delta_2^N|b_2\rangle)_{b^{N}}\\
&+\beta_2|a_2\rangle(\delta_2^1|b_1\rangle+\delta_1^1|b_2\rangle)_{b^{1}}(\delta_2^2|b_1\rangle+\delta_1^2|b_2\rangle)_{b^{2}}\cdot\cdot\cdot(\delta_2^N|b_{1}\rangle+\delta_1^N|b_2\rangle)_{b^{N}}].
\end{split}
\end{eqnarray}
If the electron spins in QD$_1$ and QD$_2$ are respectively in the state $|\downarrow\rangle_1$ and $|\downarrow\rangle_2$, with respective additional phase-flip operations $|L\rangle_a\rightarrow-|L\rangle_a$ and $|a_2\rangle\rightarrow-|a_2\rangle$ on photon $a$, the photon system consisting of one control photon and $N$ target photons is also projected into the state as shown in Eq.~(\ref{eq15}), which shows that there is a bit flip of the state on the polarization DOF for each target photon $b^n$ when the polarization of the control photon $a$ is in the state $|L\rangle_a$, and there is a bit flip of the state on the spatial-mode DOF for each target photon $b^n$ when the spatial mode of the control photon $a$ is in the state $|a_2\rangle$. Meanwhile, Eq.~(\ref{eq15}) is obtained under the consideration of nearly realistic condition, in which there are no terms show the happening of bit-flip errors. Furthermore, from Eq.~(\ref{eq14}), one can see $T^{2(N+1)}$ is an integrated coefficient for the state of the system, so the fidelity of hyperparallel photonic quantum CNOT$^N$ gate is robust to the realistic experimental parameters, which relaxes the requirement for experiment. Overall, the approach of self-error-corrected hyperparallel quantum CNOT gate for a two-photon system is successively generalized to achieve the self-error-corrected hyperparallel quantum CNOT$^N$ gate for a multiple-photon system, which eliminates bit-flip errors, guarantees a robust fidelity, and relaxes the requirement for experiment, and its failure is heralded by single-photon detectors.

\begin{figure}[ht!]
\centering
\includegraphics[width=8cm,angle=0]{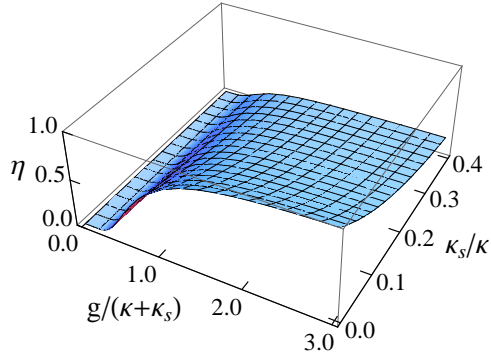}
\caption{ The efficiency of our hyperparallel photonic quantum CNOT gate.
$\omega=\omega_c=\omega_{X^{-}}$ and $\gamma/\kappa=0.1$, which are experimentally achievable, are
taken here.} \label{fig3}
\end{figure}

\section{Discussion and summary} \label{sec5}

The efficiency is defined as the ratio of the number of the output
photons to the input photons. The efficiency of the self-error-corrected hyperparallel photonic quantum CNOT gate is described as
\begin{eqnarray}\label{eq16}
\eta\;=\;|T|^8,
\end{eqnarray}
which varies with the parameter $g/(\kappa+\kappa_s)$ and $\kappa_s/
\kappa$ as shown in
Fig.~\ref{fig3}. The non-unity efficiency of our proposal originates from photon loss. For example, there is a probability that a part of the photons are detected by single-photon detectors in the two basic blocks and these photons can also lose when passing through partially transmitted mirrors. For a practical scattering condition with $g/(\kappa+\kappa_s)=3$ and $\kappa_s/
\kappa=0.1$, the
efficiency of our self-error-corrected hyperparallel photonic quantum CNOT gate can reach $65.13\%$. With the same experimental parameters, the efficiencies of our self-error-corrected hyperparallel photonic quantum CNOT$^2$ gate and CNOT$^3$ gate can reach $52.56\%$ and $42.42\%$, respectively.

The fidelities of our self-error-corrected hyperparallel photonic quantum CNOT gate and CNOT$^N$ gate are robust to the experimental parameters $g$, $\omega$, $\omega_{c}$, $\omega_{X^{-}}$, $\gamma$, $\kappa$, and $\kappa_s$, which relaxes the requirement for experiment. Exciton dephasing is a slight factor which would have a negative effect on the fidelity \cite{qdtime5}. Taking the microcavity photon lifetime $\tau=4.5ns$ and the electron spin coherence time $\Gamma_{2}\simeq2.6\mu s$ \cite{qdtime6} into account, the fidelity of every QD-microcavity system would be affected by an amount of $[1-exp(-\tau/T_2)]\simeq0.002$. Meanwhile, the
imperfect mixing of heavy-light hole would lead to an imperfect
optical selection which could affect the fidelity a little as well
\cite{qd1}. This imperfect mixing can be suppressed by engineering
the shape and the size of QDs or by choosing the types of QDs.

The accomplishment of our self-error-corrected hyperparallel photonic
quantum CNOT gate and CNOT$^N$ gate requires the technique
of the measurement on the state of the electron spin in the QD in
the orthogonal basis $[|\uparrow\rangle, |\downarrow\rangle]$, which
can be realized by measuring the helicity of the transmitted or
reflected photon. Meanwhile, two transitions
$|\uparrow\rangle\leftrightarrow|\uparrow\downarrow\Uparrow\rangle$
and
$|\downarrow\rangle\leftrightarrow|\downarrow\uparrow\Downarrow\rangle$
couple to two microcavity modes with right- and left-circular
polarization. So the microcavity for supporting two circularly polarized
modes with the same frequency is required. This requirement can be
satisfied \cite{modes1,modes2,modes3,modes4}. For example, in 2012,
Luxmoore \emph{et al.} precisely tuned the energy split between the
two circularly polarized microcavity modes to just 0.15 $nm$
\cite{modes1}. Also, the accomplishment of our scheme needs optical switches, which can couple different photons in and out the circuit of photonic quantum gate and couple different spatial modes of one photon in and out the circuit of the basic block. The performance of optical switch such as loss, delays, and the destruction on the photons, would affect the performance of our scheme. Fortunately, the suitable ultrafast optical switching device, that enables us to route single photons for quantum information processing, has been demonstrated. These switches exhibit a minimal loss, high speed performance, and a high contrast without disturbing the quantum state of photons that pass through them \cite{os1,os2,os3}. For instance, an ultrafast switch has been experimentally demonstrated and it completes the switch operation in 10$ps$ \cite{os1}, which is much shorter than the electron spin coherence time of a charged QD (3$\mu s$) \cite{qdtime1,qdtime2}. In our scheme, the failure-heralded character relies on the clicks of single-photon detectors in basic blocks. When there is a click of either single-photon detector, the process of self-error-corrected hyperparallel photonic CNOT gate and CNOT$^N$ gate  fails and it will be terminated. If the single-photon detector is perfect enough, no click of the single-photon detector marks the success of the scheme.

For establishing scalable quantum computation, multiqubit gates are useful. As well known, they can be constructed by two-qubit gates and single-qubit gates \cite{BBC95}. The direct realization of a multiqubit gate with one control qubit and multiple target qubits is essential and more efficient, and this topic  has attracted much attention \cite{atgate6,hyper5,nqubit1,nqubit2,nqubit3,nqubit4}. Fortunately, through directly generalizing our approach of self-error-corrected hyperparallel photonic quantum CNOT gate working on a two-photon system, the self-error-corrected hyperparallel photonic quantum CNOT$^{N}$ gate working on a multiple-photon system is achieved, in which one photon in the polarization and the spatial-mode DOFs respectively controls all of the $N$ target photons in the polarization and the spatial-mode DOFs. Our self-error-corrected hyperparallel photonic quantum CNOT$^N$ gate is directly achieved rather than constructed by universal quantum gates, which has a potential in performing hyperentanglement preparation \cite{SB01}, error corrections \cite{TCT14}, and quantum algorithms \cite{BR01,LFL15}.

In summary, we have presented the first scheme for designing the self-error-corrected hyperparallel
photonic quantum CNOT gate, which relaxes the problem of the bit-flip errors happening in the quantum gate in the nearly realistic condition. The self-error-corrected pattern of our hyperparallel photonic quantum CNOT gate decreases the bit-flip errors, guarantees the robust fidelity, and relaxes the requirement of its experiment. Meanwhile, our self-error-corrected hyperparallel photonic quantum CNOT gate works in a failure-heralded way, as the clicks of single-photon detectors can mark the failure of the self-error-corrected hyperparallel photonic quantum CONT gate. Moreover, it can be generalized to the self-error-corrected hyperparallel photonic quantum CONT$^N$ gate, which also works in the failure-heralded way. These features make our scheme more useful in quantum computation and the multi-photon hyperentangled-state generation in quantum communication in the future.

\section*{ACKNOWLEDGEMENTS}

This work is supported by the National Natural Science Foundation of
China under Grants No. 11474026, No. 11674033, and No. 11505007, and
the Fundamental Research Funds for the Central Universities under
Grant No. 2015KJJCA01.

%National Natural Science Foundation of China (NSFC) (11474026, 11674033, 11505007); Fundamental Research Funds for the
%Central Universities (2015KJJCA01).

%\end{CJK*} %显示中文

\end{document}